\documentclass[a4paper,twoside]{article}
\usepackage[authoryear]{natbib}
\bibpunct{(}{)}{;}{a}{}{,}
\usepackage{graphicx}
\date{} 
\newcommand{\nbody}{$N$-body}
\newcommand{\Eq}[1]{Eq.~(\ref{#1})}
\newcommand{\Fig}[1]{Fig.~\ref{#1}}
\newcommand{\Sec}[1]{Section~\ref{#1}}

\newcommand{\mlapm}{\texttt{MLAPM}}
\def\ea{et~al.~}                            
\def\LCDM{$\Lambda$CDM}

\def\hMpc{$h^{-1}{\ }{\rm Mpc}$}
\def\hMsun{$h^{-1}{\ }{\rm M_{\odot}}$}
\def\nbody{$N$-body}
\def\c15{$c_{\rm 1/5}$}

\def\fracj#1#2{{\textstyle{#1\over#2}}}

\def\d{{\rm d}}

\def\i{\relax\ifmmode{\rm i}\else\char16\fi}

\title{\large\bf\flushleft How to simulate the Universe in a Computer}
\author{\parbox{\textwidth}{\flushleft
\vspace{-0.5cm}
{\it Alexander Knebe$^{1,2}$\\
\vspace{0.4cm}
{\small $^{1}$Centre for Astrophysics \& Supercomputing, Swinburne University, PO Box 218, Mail \#31, Hawthorn, VIC 3122, Australia\\
        $^{2}$Astrophysikalisches Institut Potsdam (AIP), An der Sternwarte 16, 14482 Postdam, Germany\\      
  Email: aknebe@aip.de}}}}
\begin{document}
\twocolumn[
\begin{changemargin}{.8cm}{.5cm}
\begin{minipage}{.9\textwidth}
\vspace{-1cm}
\maketitle
%
%
\small{\bf Abstract:}

In this contribution a broad overview of the methodologies of
cosmological \nbody\ simulations and a short introduction explaining
the general idea behind such simulations is presented.  After
explaining how to set up the initial conditions using a set of $N$
particles two (diverse) techniques are presented for evolving these
particles forward in time under the influence of their
self-gravity. One technique (tree codes) is solely based upon a
sophistication of the direct particle-particle summation whereas the
other method relies on the continuous (de-)construction of arbitrarily
shaped grids and is realized in adaptive mesh refinement codes.



\medskip
\medskip
\end{minipage}
\end{changemargin}
]
\small

\section{Introduction}

The purpose of cosmological simulations is to model the growth of
structures in the Universe. They have a long history and numerous
applications. These simulations play a very significant role in
cosmology because they can be considered as an ``experiment'' to
verify theories of the origin and evolution of the Universe.

The Universe is believed to have started with a Big Bang in which --
or more precisely: shortly after which -- tiny fluctuations (in an
otherwise homogeneous and isotropic space) were imprinted into the
radiation and matter density field. To understand how the Universe
evolved from that early stage into what we observe today (i.e. stars,
galaxies, galaxy clusters, ...) one needs to follow the evolution of
those density fields using numerical methods as soon as they turn
non-linear.  Therefore, the approach to cosmological simulations is
actually twofold: firstly, one needs to generate the initial
conditions according to the cosmological structure formation model to
be investigated and secondly, the initial density field (sampled by
the particles) needs to be evolved forward in time using an \nbody\
code.

In all such codes the evolution is simulated by following the
trajectories of particles under their mutual gravity.  These particles
are supposed to sample the matter density field as accurately as
possible and a cosmological simulation is nothing more (and nothing
less) than a simple and effective tool for investigating non-linear
gravitational evolution.  There are two constraints on a cosmological
simulation though: a) the correct initial conditions and b) the
observation of galaxies, galaxy clusters, large-scale structure,
voids, etc. Simulations are hence trying to bridge the gap between
observations of the early Universe (i.e. anisotropies in the Cosmic
Microwave Background observed as early as 300000 years after the Big
Bang) and the Universe as we see it today.

The first application of \nbody\ methods in astrophysics was in the
simulations of star clusters using as little as a handful of
particles (Aarseth 1970). During the 1970's more simulations of galaxy
clustering were performed using what is called PP
(\textbf{p}article-\textbf{p}article) methods (Peebles 1970) and not
earlier than 1981 the first cosmological simulations using more than
20000 particles became feasible (Efstathiou~\ea 1981).

Until now the methods have been continuously refined to allow for more
and more particles while simultaneously resolving finer and finer
structures. Today it is standard to run a cosmological simulation with
millions of particles in a couple of days on large supercomputers or
even clusters of PC's (cf. Gill, Knebe~\& Gibson 2004). These
simulations can resolve the orbits of satellite galaxies within dark
matter haloes spanning five orders of magnitude in mass and a spatial
dynamical range well above 30000.

In this contribution I would like to focus on two numerical techniques
in particular presenting their methodologies, advantages and
shortcomings when being compared with each other. I need to stress
though that all cosmological simulations are based on the assumption
that the Universe mainly consists of dark matter interacting merely
gravitationally. Baryonic matter, which only accounts for about 15\%
of the total mass, is accounted for in such simulations only via its
gravitational effects, too. Even though todays simulation methods and
computer technology have become sufficiently sophisticated as to allow
for hydrodynamical processes to be included, the particulars of such
implementations, however, are beyond the scope of this contribution.

\begin{figure*}[h]
\begin{center}
\includegraphics[scale=0.80, angle=0]{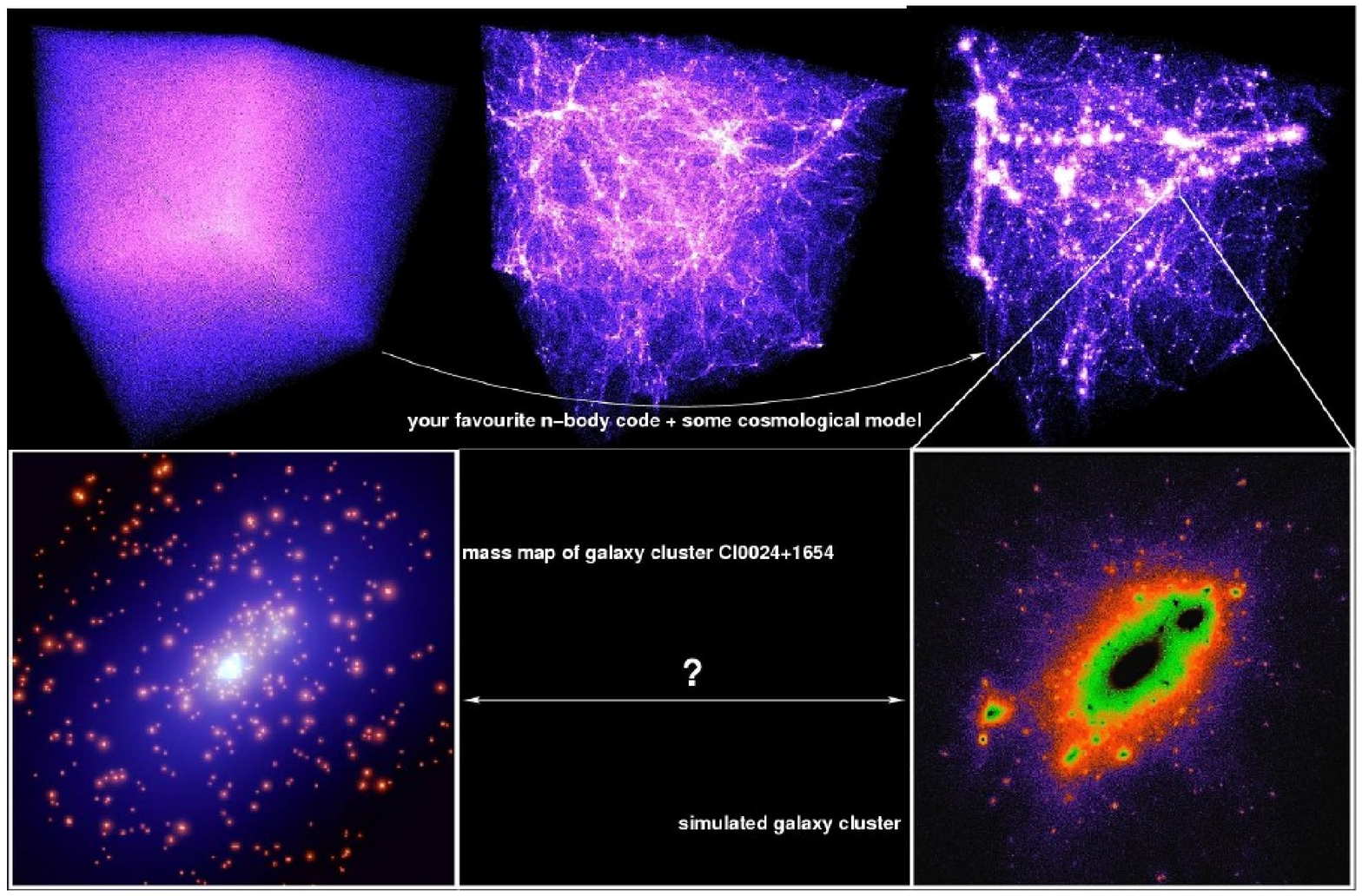}
\caption{Illustration of the idea driving \nbody\ simulations. Initial conditions
         are being evolved forward in time modeling gravity alone. The
         outputs over time are then compared to observational data and the
         cosmological model adjusted accordingly. 
         Image credit (mass map of Cl0024+164): European Space Agency, 
         NASA and Jean-Paul Kneib (Observatoire Midi-Pyrenees, 
         France/Caltech, USA)}\label{Nbody}
\end{center}
\end{figure*}

\Fig{Nbody} depicts the conceptual ideas behind
(cosmological) \nbody\ simulations: starting from initial seed
inhomogeneities superimposed onto a homogeneous and isotropic
background the matter field is evolved forward in time. This evolution
depends on the cosmological model under investigation and is performed
using an \nbody\ code. Snapshots of the simulation at various times are
recorded and then analysed and compared to observational data to
verify and falsify theories of structure formation and evolution.

\begin{figure*}[h]
\begin{center}
\includegraphics[scale=0.65, angle=0]{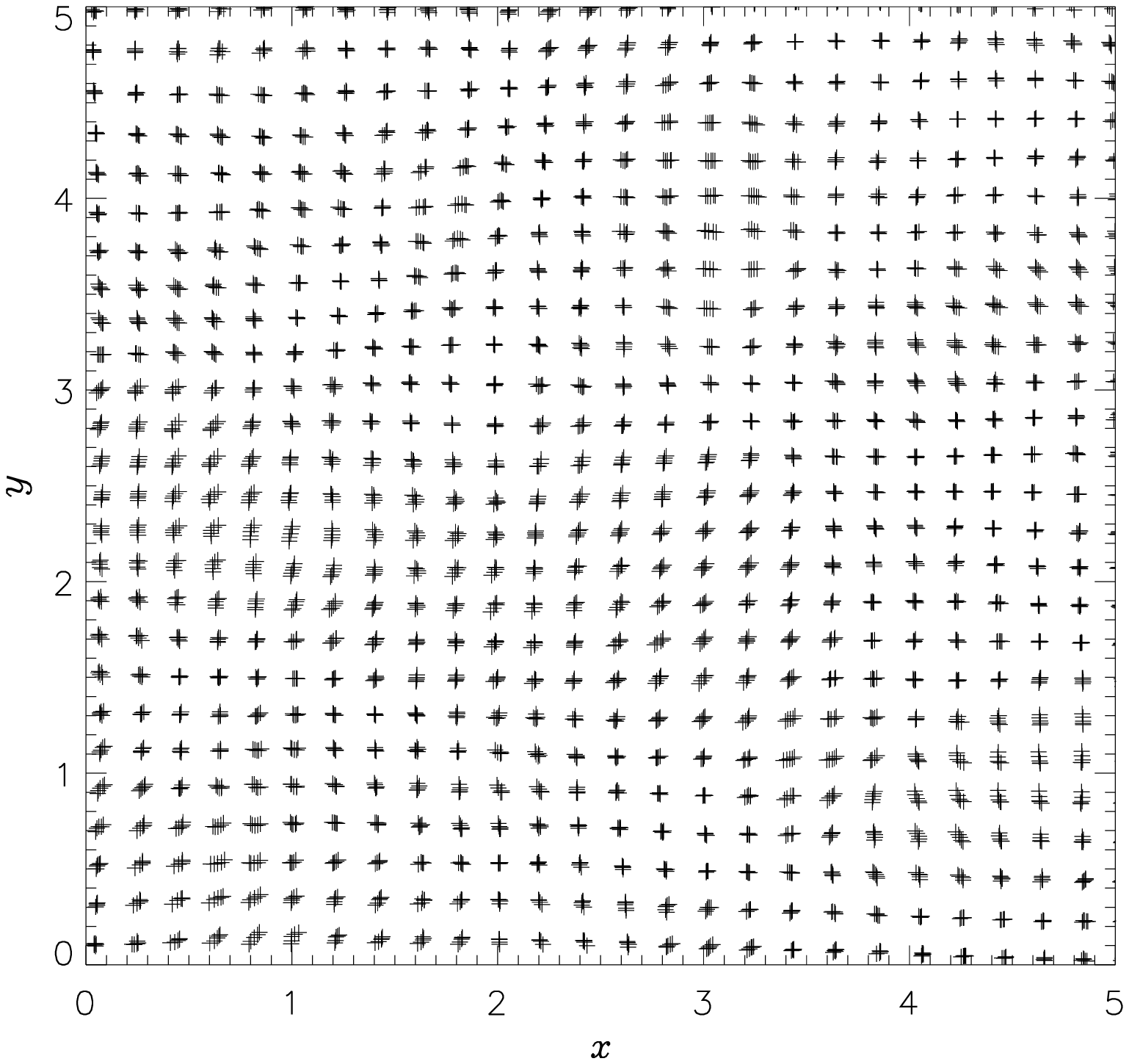}
\caption{Initial conditions according to the Zeldovich approximation 
        (cf. \Eq{zeldovich}). One clearly sees the decomposition of 
        density perturbations as waves.}\label{IC}
\end{center}
\end{figure*}

\section{{The Initial Conditions}}
The most common way to set up initial conditions for a cosmological
simulation is to make use of the Zeldovich approximation to move
particles from a Lagrangian point $\vec{q}$ to a Eulerian point
$\vec{x}$ (e.g. Efstathiou~\ea 1985):

\begin{equation} \label{zeldovich}
 \vec{x} = \vec{q} - D(t) \vec{S}(\vec{q}) \ ,
\end{equation}

\noindent
where $D(t)$ describes the growing mode of linear fluctuations and
$\vec{S}(\vec{q})$ is the 'displacement field'. The initial Lagrangian
coordinates $\vec{q}$ are usually chosen to form a regular,
three-dimensional lattice.

The displacement field $\vec{S}(\vec{q})$ is related to a
pre-calculated input power spectrum of density fluctuations, $P(k)$,
which in turn depends on the cosmological model under consideration

\begin{equation} \label{FourierSum}
 \vec{S}(\vec{q}) = \nabla_q \Phi(\vec{q}), \ \ \
 \Phi(\vec{q})    = \sum_{\vec{k}} a_{\vec{k}} \cos(\vec{k} \cdot \vec{q}) + 
                                   b_{\vec{k}} \sin(\vec{k} \cdot \vec{q}) \ ,
\end{equation}

\noindent
where the Fourier coefficients $a_{\vec{k}}$ and $b_{\vec{k}}$ are
linked to $P(k)$ and are given as

\begin{equation} \label{FourierAmplitudes}
 a_{\vec{k}} = R_1 \frac{1}{k^2} \sqrt{P(k)} , \ \ \ 
 b_{\vec{k}} = R_2 \frac{1}{k^2} \sqrt{P(k)} .
\end{equation}

\noindent
$R_1$, $R_2$ are (Gaussian) random numbers with zero mean and
dispersion unity.

An example for the Zeldovich approximation ``at work'' can be found in
\Fig{IC} where a slice through the initial conditions for a standard
\LCDM\ simulations at redshift $z=50$ is shown. This figure nicely
demonstrate the decomposition of the density perturbations as waves.

\section{{Solving Poisson's Equation}}

When used to model the dynamics of a collisionless system such as dark
matter, an $N$-body code aims at simultaneously solving the collisionless
Boltzmann equation (CBE) 

\begin{equation} \label{CBE}
 \frac{\partial f}{\partial t} + \sum_{i=1}^3 
 \left( 
       v_i\frac{\partial f}{\partial x_i} - \frac{\partial \Phi}{\partial x_i} \frac{\partial f}{\partial v_i} = 0
 \right)
\end{equation}

\noindent
and Poisson's equation

\begin{equation} \label{poisson}
 \nabla^2 \Phi (\vec{r}) = 4 \pi G \rho (\vec{r})\ .
\end{equation}

\noindent
The CBE \Eq{CBE} is solved by the method of characteristics (e.g.,
Leeuwin, Combes \& Binney 1993). Since the CBE states that $f$ is
constant along any trajectory $\{ \vec{r}(t), \vec{v}(t)\}$, the
trajectories obtained by time integration of $N$ points
$\{\vec{r}_i,\vec{v}_i \}$ sampled from the distribution function $f$
at time $t=t_{\rm initial}$ form a representative sample of $f$ at
each time $t$.

Hence the problem reduces to solving Poisson's \Eq{poisson} for a set
of $N$ particles and advancing them forward in time according to the
equations of motion derived from the system's Hamiltonian $H$
(remember that \Eq{CBE} can be written as $\partial f/\partial t +
[f,H]=0$). The details of the time-integration of the equations of
motions are going to be explained later on in
\Sec{Newton} though; in this Section I like to focus on the gravity
solver.

\noindent
Currently there are two commonly used approaches for deriving the
potential from Poisson's equation: a) tree codes rely on a direct
particle-particle summation, and b) PM
(\textbf{p}article-\textbf{m}esh) codes utilize a numerical
integration of \Eq{poisson} on a grid.

\begin{figure*}[h]
\begin{center}
\includegraphics[scale=0.5, angle=0]{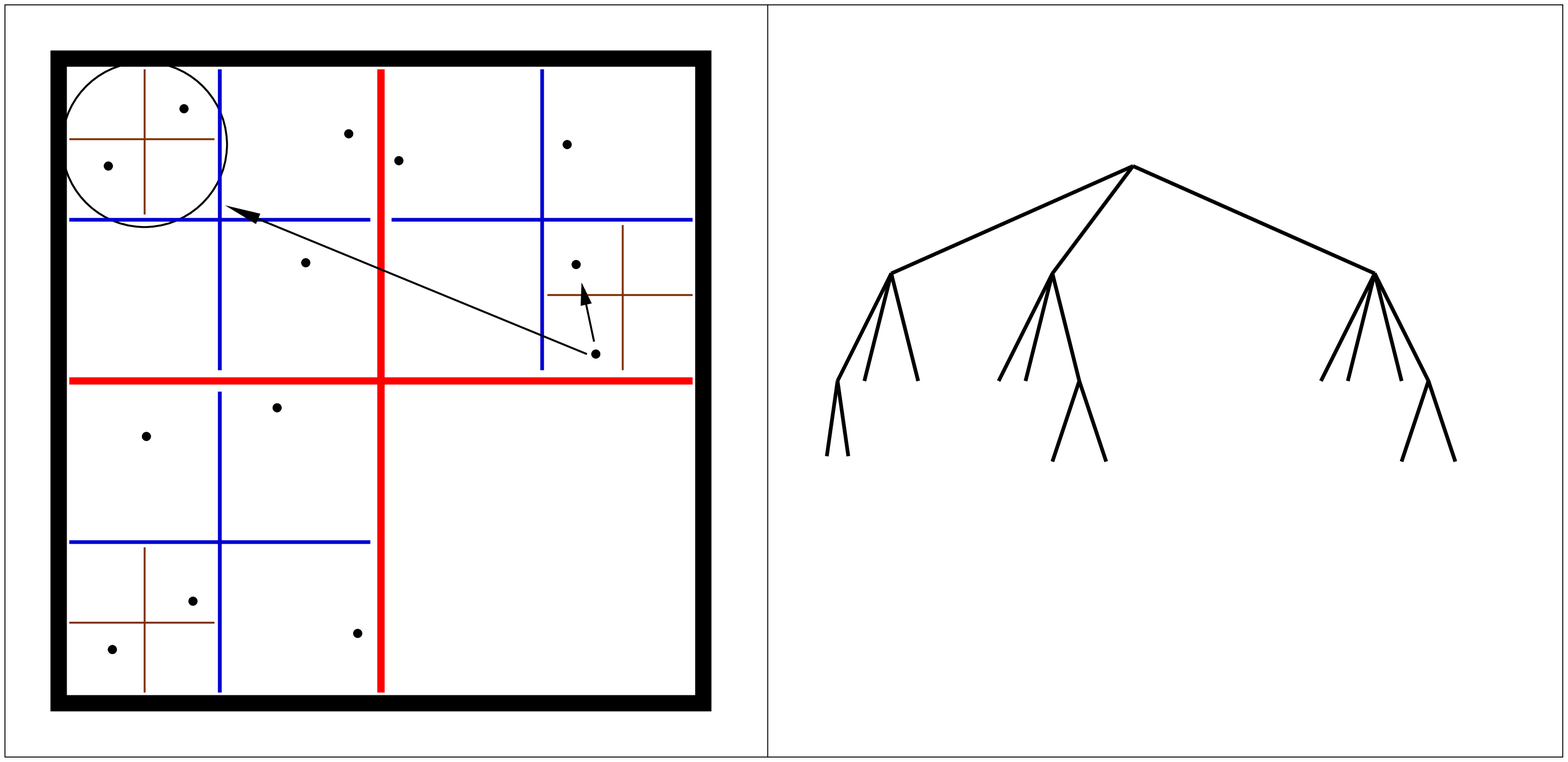}
\caption{Illustration of a tree code. The left panel shows the actual particle
         distribution and its cubical decomposition. The right panel is the
         tree corresponding to this distribution.}\label{AdaptiveParticles}
\end{center}
\end{figure*}

\subsection{{Tree Codes}} 

\subsubsection{The Pre-Requisites}
The \textbf{p}article-\textbf{p}article (PP) method upon which tree
codes are based assumes that the particles are $\delta$-functions and
hence the density field (rhs of Poisson's equation~\Eq{poisson}) reads
as follows:

\begin{equation}\label{delta}
\rho(\vec{r}) = \sum_{i=1}^{N} m_i \delta (\vec{r}-\vec{r}_i) \ ,
\end{equation}

\noindent
where $N$ is the total number of particles in use.

\subsubsection{The Forces}
Combining \Eq{delta} with \Eq{poisson} the analytical solution for the
force $\vec{F}$ at particle position $\vec{r}_i$ is given by:

\begin{equation}\label{PPforce}
\vec{F}(\vec{r}_i) = \sum_{j\ne i} \frac{m_i m_j}{|\vec{r_i}-\vec{r_j}|^2} \ .
\end{equation}

\noindent
But as we are interested in deriving the force at every single
particle position, the PP method scales like $N^2$ ($N$ summations,
each over $(N-1)$ particles). Therefore, a (straightforward) PP summation
appears not to be feasible for evolving a set of $N$ particles under
their mutual gravity, not even on the largest supercomputers available
nowadays! One needs to bypass the increase in computational time for
large numbers of particles with a more sophisticated treatment when
calculating the forces. One way of achieving this is to organize the
particles in a tree-like structure: particles located ''far away''
from the actual particle (at which position we intend to calculate the
force) can be lumped together as a single -- but more massive --
particle. This tunes down the number of calculations dramatically.

The idea of a tree code is sketched in \Fig{AdaptiveParticles}.  The
particles are organized in a tree-like structure based upon a cubical
decomposition of the computational domain.  Consequentially, for each
particle we ``walk the tree'' and add the forces from branchings that
need no further unfolding into finer branches according some
pre-selected ``opening criterion''.

One publicly available tree code is called GADGET\footnote{GADGET can
be downloaded from at this web address
\texttt{http://www.mpa-garching.mpg.de/gadget}} (\textbf{Ga}laxies
with \textbf{D}ark Matter and \textbf{G}as int\textbf{E}rac\textbf{T})
and I refer the reader to a more elaborate discussion of this
technique to its reference paper by Springel, Yoshida~\& White (2001).

\subsubsection{Force Resolution}\label{epsilon}
In order to avoid the singularity for $\vec{r}_i=\vec{r_j}$ in
\Eq{PPforce} one needs to set a limit on the minimal
allowed spatial separation between two particles. This can be achieved
by introducing a (fixed) scale, i.e. the softening
parameter~$\epsilon$:

\begin{equation}\label{softening}
\vec{F}(\vec{r}_i) = \sum_{j\ne i} \frac{m_i m_j}{|\vec{r_i}-\vec{r_j}|^2 + \epsilon^2}
\end{equation}

This softening is closely related to the overall force resolution of
the simulation and an elaborate discussion of it can be found in
Dehnen (2001).

\subsection{{Particle-Mesh Codes}} 

\subsubsection{The Pre-Requisites} \label{PMrequisites}
Another way for obtaining the forces is to numerically integrate
Poisson's equation~(\ref{poisson}). This method, however, demands the
introduction of a grid in order to define the density and hence the
name
\textbf{p}article-\textbf{m}esh (PM) method.  The grid is usually of a
regular (cubic) shape with $L \times L \times L$ cells where each cell
is identified by the index triplet $(i,j,k)$.  The forces are then
calculated according to the following scheme:

\begin{enumerate}
 \item[1.] assign all particles to the grid to get $\rho_{i,j,k}$
 \item[2.] solve Poisson's equation $\nabla^2 \phi_{i,j,k} = 4 \pi G \rho_{i,j,k}$
 \item[3.] differentiate to get forces $F_{i,j,k} = - \nabla \phi_{i,j,k}$
 \item[4.] interpolate $F_{i,j,k}$ back to particle positions
\end{enumerate}

\subsubsection{The Forces}
With this scheme most of the time is spent in step 2 and the most
common way to solve Poisson's equation on a grid is to make use of
FFT's (\textbf{F}ast-\textbf{F}ourier-\textbf{T}ransforms).  The
analytical solution to Poisson's equation is given by the integral

\begin{equation}\label{green}
 \Phi (\vec{r}) = \int G(\vec{r}-\vec{r}') \rho(\vec{r}') d\vec{r}'
\end{equation}

\noindent
where $G(\vec{x}) = -\vec{x}/x^{3/2}$ is the Green's function of
Poisson's equation. This integral can readily be evaluated in
Fourier-space, i.e.

\begin{equation}
  \hat{\Phi} = \hat{G} \  \hat{\rho}
\end{equation}

\noindent
where $\hat{\Phi}$, $\hat{G}$, and $\hat{\rho}$ are the Fourier
transforms of the respective variables.

The PM approach proves to be exceptionally fast outperforming any tree
code.

There are of course other techniques than the use of FFT's available
to numerically solve Poisson's equation but the utilisation of FFT's
is the most common approach as it appears to be the fastest.

\subsubsection{Force Resolution}
The most severe problem with the PM method is the lack of spatial
resolution below two grid spacings. Whereas tree codes require the
introduction of a softening length to avoid the force singularity for
close encounters of particles PM codes suffer from the opposite
problem. Gravity is an attractive force and hence the particles flow
from low density regions into high density regions amplifying
primordial density fluctuations.  This leads to an excess of particles
in certain cells whereas other cells are becoming more and more devoid
of matter. But as the spacing of the grid introduces a (smoothing)
scale particles closer than about two cell distances do not interact
according to \Eq{green} anymore. While we had to introduce a force
softening for tree codes to avoid two-body interactions and make the
simulation collisionless, respectively, the PM method naturally admits
such a smoothing. However, we are left with the situation where we can
not resolve structure formation on scales of (and below) roughly the
cell spacing of the grid!

This is a major problem and the most obvious way to overcome it is to
introduce finer grids in regions of high density. These grids though
need to freely adapt to the actual particle distribution at all times
and hence navigating such complex grids through computer memory is a
very demanding task.  One of the freely available \textbf{a}daptive
\textbf{m}esh \textbf{r}efinement (AMR) codes is
\mlapm\footnote{\mlapm\ can be downloaded from this web address
\texttt{http://www.aip.de/People/AKnebe/MLAPM}} (Knebe, Green~\& Binney
2001).

\begin{figure*}[h]
\begin{center}
\includegraphics[scale=0.7, angle=0]{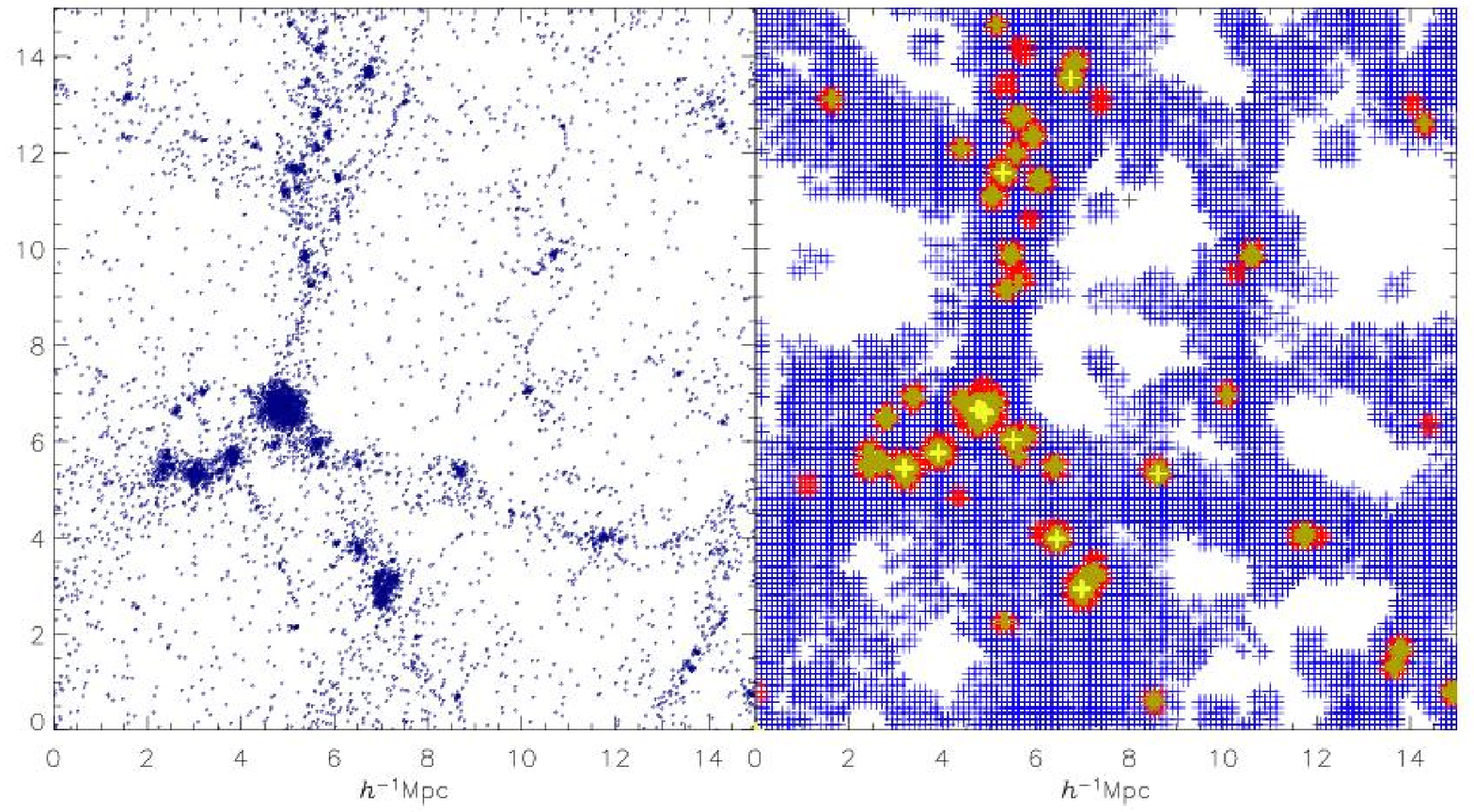}
\caption{An example for adaptive mesh refinement. The left panel
         shows the particle distribution at redshift $z=0$ in a
         \LCDM\ simulation. The right panel indicates the arbitrarily
         shaped grids invoked by the AMR code \mlapm\ to solve
         Poisson's equation.}\label{MLAPMref}
\end{center}
\end{figure*}

The mode of operation of this AMR technique can be viewed in
\Fig{MLAPMref} where a slice through a standard \LCDM\ simulation is
presented. The left panel shows the distribution of particles whereas
the right panel indicates the adaptive meshes used to obtain a
solution to Poisson's equation.

It needs to be stressed though that the use of irregularly shaped grids
inhibits FFT's. Another technique for solving Poisson's equation
needs to be sought such as, for instance, multi-grid relaxation
(cf. Knebe, Green~\& Binney 2001).

\subsection{Hybrid Methods}
There are, of course, various other techniques for simultaneously
being time efficient and having a credible force resolution.  One
possibility is realized in the so-called P$^3$M technique
(i.e. Couchman 1991) where a combination of PP and PM provides the
necessary balance between accuracy and efficiency: the force as given
by the plain PM calculation is augmented by a direct summation over
all neighboring particles within the surrounding cells. This gives
accurate forces down to the scale provided by the softening parameter
$\epsilon$ again. Other examples, for instance, are Tree-PM (Bode~\&
Ostriker 2003) and moving mesh (Gnedin 1995) codes, but the details
are well beyond the scope of this contribution.

\subsection{Mass Resolution}
It still needs to be mentioned that a cosmological simulation in
practice only simulates a certain fraction of the Universe. This is
what people refer to as the \textit{simulation box}. However, to
account for the fact the Universe is actually infinite one uses
periodic boundary conditions: particles leaving the box on one side
immediately enter the box again on the other side.

Moreover, the size of the box also defines the mass resolution of the
simulation. We are only using a certain number of particles within a
fixed region of the Universe. And as the density of the Universe is
determined by the cosmological model under investigation, each
individual particle has a certain mass. This mass determines the
\textit{mass resolution} of that specific simulation. For instance, if
we model the evolution of about 2 million particles in a box with side
length 25\hMpc\ using the \LCDM\ cosmology ($\Omega_0 = 0.3$), each
particle weighs about $6
\cdot 10^{8}$\hMsun. Therefore we will not be able to resolve dwarf galaxies in
that particular cosmological simulation ($M_{\rm dwarf} \geq
10^7$\hMsun).

\subsection{{Comparison}}
It only appears natural to ask the question which method is superior
and how they compare, respectively.

There is no straight forward answer as both methods have their
(dis-)advantages. Tree codes are based upon the assumption that the
Universe is filled with particles of a certain size related to the
softening $\epsilon$ (cf. \Sec{epsilon}). Adaptive mesh refinement
codes use a smoothed density field (which in turn also introduces an
effective particle size) to obtain the potential and hence the force
field. In both cases it is important to bear in mind that particles
are only to be understood as markers in phase-space and should not
interact on a two-body basis, i.e. one always intends to integrate the
collisionless Boltzmann equation~(\ref{CBE}). There are several
studies investigating two-body interactions in such simulations and it
can be confirmed that they are more prominent in tree codes (Binney~\&
Knebe 2002). But as long as one complies with certain constraints on
the numerical parameters (cf. Power~\ea 2003) such effects can be
minimized.

There are several studies comparing tree and AMR codes both in
efficiency and accuracy (cf. Frenck~\ea 1999, Knebe~\ea 2001,
O'Shea~\ea 2003) but in the end it all comes down to a ``question of
taste''. Both techniques are well enough developed to successfully
model the formation and evolution of cosmic structures.

\section{{Newtonian Mechanics in an Expanding Universe}} \label{Newton}

Even though solving Poisson's equation is the heart and soul of every
\nbody\ code, it is also important to accurately update particle
positions and velocities, i.e. integrating the equations of motion.
And as the Universe is expanding it is convenient to introduce
\textit{comoving coordinates}:

\begin{equation}\label{comoving}
 \vec{x} = \vec{r} / a(t) \,
\end{equation}

\noindent
where $a(t)$ is the cosmic expansion factor.

\noindent
The (comoving) Lagrangian is given by

\begin{equation}
 {\mathcal L} = \fracj12 a^2 \dot{x}^2 - { \Phi\over a},
\end{equation}

\noindent
which leads to the canonical momentum

\begin{equation}
\vec{p} = a^2 \dot{\vec{x}} .
\end{equation}

\noindent
Hamilton's equations are therefore

\begin{equation}\label{He}
 \begin{array}{lcc}
  \displaystyle \frac{\d \vec{x}}{\d t} & = &\displaystyle  {\vec{p}\over a^2} \\ 
\\
  \displaystyle \frac{\d \vec{p}}{\d t} & = &\displaystyle  -{\nabla \Phi\over a} \ .
 \end{array}
\end{equation}

\noindent
These equations can be discretized and integrated using a second-order
accurate scheme as follows:

\begin{eqnarray}\label{stepping}
  \vec{x}_{n+1/2} & = & \vec{x}_n + \vec{p}_n
                 \int_t^{t+\Delta t/2}{\d t\over a^2} \nonumber\\
  \vec{p}_{n+1}  & = & p_n - \nabla\Phi(x_{n+1/2})\int_t^{t+\Delta t}
                 {\d t\over a}\\
  \vec{x}_{n+1}  & = & \vec{x}_{n+1/2} + \vec{p}_{n+1}
              \int_{t+\Delta t/2}^{t+\Delta t}{\d t \over a^2},  \nonumber
\end{eqnarray}

\noindent
where the integrals can be evaluated analytically as they depend only
on the cosmology.

This modified \textit{leap-frog scheme} only needs to store one copy
of the positions and velocities whereas other integrators as, for
instance, Runge-Kutta consume more memory.

\section{{Final Remarks}}
A broad overview of the concepts behind cosmological \nbody\
simulations has been presented. After a short introduction explaining
the needs for such simulations I explained how to actually set up the
initial conditions. There are numerous techniques for evolving that
set of $N$ particles forward in time under the influence of its own
self-gravity alone, but I focused on two (diverse) methods, namely
tree codes and adaptive mesh refinement codes. Their mode of operation
has been introduced and their limitations pointed out and compared with
each other.

This contribution can only be understood as a very brief and general
introduction to the ideas behind cosmological \nbody\ simulations.  It
is far from being complete and exhaustive and I refer the reader to
more elaborate review articles such as Bertschinger (1998) and the
monograph ``Computer Simulations using Particles'' by Hockney~\&
Eastwood (1988).

\section*{Acknowledgments} 
I like to thank the organizers of the ``Gravity 2004'' workshop,
especially Stuart P.D. Gill and Geraint Lewis.


\end{document}